**Reorganization of a 2D disordered granular medium due to a small local cyclic perturbation**

E. Kolb[1], C. Goldenberg[2], S. Inagaki[1], E. Clément[1]

[1] *ESPCI / UPMC, PMMH (Laboratoire de Physique et Mécanique des Milieux Hétérogènes), Equipe Granulaires, 10 rue Vauquelin, 75231 Paris Cédex 05 France,* kolb@ccr.jussieu.fr
[2] *Laboratoire de Physique de la Matiére Condensée et des Nanostructures, Université Lyon 1; CNRS, UMR 5586, Domaine Scientifique de la Doua, F-69622 Villeurbanne cedex, France.*

**Abstract**

We measure experimentally the rearrangements due to a small localized cyclic displacement applied to a packing of rigid grains under gravity in a 2D geometry. We analyze the evolution of the response to this perturbation by considering the individual particle displacement and the coarse grained displacement field, as well as the mean packing fraction and coordination number. We find that the displacement response is rather long ranged, and evolves considerably with the number of cycles. We show that a small difference in the preparation method (induced by tapping the container) leads to a significant modification in the response though the packing fraction changes are minute. Not only the initial response but also its further evolution change with preparation, demonstrating that the system still retains a memory of the initial preparation after many cycles. Nevertheless, after a sufficient number of cycles, the displacement response for both preparation methods converges to a nearly radial field with a *1/r* decay from the perturbation source. The observed differences between the preparation methods seem to be related to the changes in the coordination number (which is more sensitive to the evolution of the packing than the packing fraction). Specifically, it may be understood as an effect of the breaking of local arches, which affects the lateral transmission of forces.

**Introduction**

The rheology (flow behavior) of dense heterogeneous materials such as nanocomposites [1], glasses [2], foams [3, 4], pastes [5] and granular materials [6] is of considerable technological and scientific interest. Due to the heterogeneity of these materials, the transmission of forces is not uniform and it is quite hard to predict the evolution of material structure when subjected to macroscopic deformation (e.g., applied shear) which may lead to complex features such as the formation of shear bands, the self-organized alignment of particles, and separation of phases of different packing fractions. Understanding the coupling between structure and dynamics in heterogeneous dense phases is necessary for the prediction of the flow properties of the material and poses a great challenge. The combined effects of steric exclusion between particles and inhomogeneities of the material lead to non-affine particle displacements (see, e.g.[2]), which cannot be simply inferred from the macroscopic applied strain field [7].

Granular materials, being composed of macroscopic grains (which do not exhibit Brownian motion), offer the possibility to visualize the dynamics in jammed phases by tracking displacements of individual grains subject to relatively simple interactions. Considerable theoretical work and simulations have been devoted to understanding the dynamics in granular materials and making some connections with other dense systems, especially for describing the concept of jamming (e.g., [8]). However, only a few experimental studies on granular systems have been performed, although these are crucial for understanding the relation between particle mobilities and the structure, as well as the rheology of dense phases ([9-14]).

The present work is focused on a two dimensional (2D) dense, disordered granular material composed of rigid grains. We describe in more detail a conceptually simple experiment [13] aimed at understanding the relations between mobilities and local structure. Instead of applying a macroscopic perturbation to the container holding the granular material(i.e., at the boundaries, as in [12]), we apply a small, cyclic perturbation directly in the bulk by moving one of the grains (the "intruder"). We test the local response of the material to this perturbation by studying the displacements of grains. From a practical point of view, this experiment is similar to standard penetrometry tests currently used in soil mechanics. However, our focus here is on the microscopic rearrangements introduced by the perturbation, which are a consequence of the fragile nature of granular material. The network of contacts between grains is not permanent, and the perturbation induced by the displacement of the intruder can open or close some contacts and produce irreversible rearrangements. Such modifications in the structure may affect the mechanical properties of the material at a larger scale (see, e.g., [15]). In order to characterize the rearrangements induced by the displacement of the intruder and their evolution with the cyclic perturbation, we measure the displacements of individual particles. We note that unlike the experiments of [12], in which only the irreversible part of the displacement was measured (i.e., between two consecutive full cycles), we also measure the total displacement (i.e., at the middle of the cycle). Some of the results obtained using this experimental method have been presented in [13]. Here we present a more systematic analysis of the macroscopic response of the system and the evolution of the mean coordination number, as well as the effect of changing the method of preparation of the system.

**Experimental set-up and procedures**

We investigate a 2D granular medium, i.e., the motion of the grains is confined to a plane. The 2D geometry allows us to follow the displacement of each grain by simply using a camera mounted above the experimental setup. The grains are hollow metallic cylinders whose form is adapted to the 2D geometry: the axes of the cylinders are perpendicular to the plane of motion (see Figure 1, right). The cylinders have two different outer diameters $d_1 = 4$ mm and $d_2 = 5$ mm and a height of 3 mm. Mixing two types of grains leads to a disordered granular medium by avoiding crystallization, i.e., a regular stacking of the grains.

Around 4000 such grains in an equal proportion in mass of the two types of cylinders (leading to a ratio of 7:4 between the numbers of small and large grains) are piled onto an inclinable plane (Figure 1, left), made of a low friction glass plate allowing a backward illumination. The lateral and bottom walls, made of Plexiglas, delimit a rectangular frame of L=26.8 cm (~ 54 $d_2$) width and an adjustable height of typically H=34.4 cm (~ 70 $d_2$). The cylinders are made of brass with a coating of nickel alloy, so that the grains are quite rigid: the deformation at the contacts is less than a hundred times smaller than the minimum displacement we can detect. Consequently, the particle displacements measured in the experiment are due to reorganizations of grains (contact opening or closing [16], relative sliding…) rather than the deformation at the contacts. The 2D packing fraction $c$ is defined as the ratio of the area of grains to the total area they occupy.

In preparing the system, the bottom plane is tilted at an angle φ (see Figure 1, right - inset). This way we control the effective gravity $g\sin\varphi$ (where $g$ is the gravitational acceleration) and thus the confining pressure inside the granular material. A value of φ =33° was chosen, larger than the static Coulomb angle of friction between the grains and the glass plate, which is around $\theta = 20°$ ($\mu_{grain/glass} = \tan\theta$ is the coefficient of static friction between the grains and the glass). Grains can therefore freely slide downward to occupy any free areas.

We used two methods of preparation: the "normal" method, in which the grains are mixed at random in a specified area [13], and a "tapping" method consisting of the same initial procedure followed by tapping the walls of the inclined container just before starting the experiment. The initial mean packing fraction is $c = 0.790 \pm 0.003$ for the tapping preparation compared with $c = 0.787 \pm 0.004$ for the normal preparation, where the error bar is the standard deviation over the 16 realizations. Evidently, the difference in preparation does not change the packing fraction significantly: $\Delta c / c = +0.3\%$, i.e., the difference in packing fraction is within the experimental error. Note that the packing fraction is calculated by counting the number of grains whose centers are located in a rectangular sub-frame at a distance of 3 $d_2$ above the intruder and 1.5 $d_2$ from the left, top and right boundaries of the camera frame.

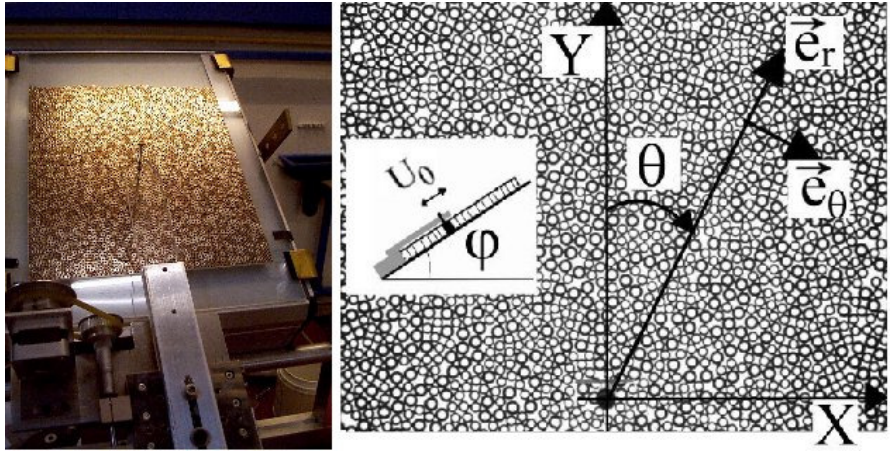

*Figure 1*. Left: Experimental setup. Right: Typical frame of observation of the packing. The intruder is below the black point (inset: sketch of the experimental setup viewed from the side). $U_0$ is the amplitude of displacement of the intruder.

The intruder is a large grain of diameter $d_2$ located in the horizontal centre of the container, 21.2 cm (i.e. ~ 42 $d_2$) below the upper free surface. The intruder is attached to a rigid Plexiglas arm (reinforced by metallic parts) moved by a translation stage and a stepping motor driven by a computer. The arm motion takes place along the vertical axis $Y$ of the container, parallel to the plane. In this report we use an intruder displacement, $U_0$, which is a fraction of a grain diameter ($U_0$ =1.25 mm = $d_2/4$) giving a mean strain in the packing of slightly less than $10^{-2}$, far above the elastic limit [17] but also far below the regime of fully developed plastic flow [18], in which shear bands typically appear. The intruder is repeatedly moved up and then back down to its initial position with the same displacement amplitude $U_0$. A total of 261 cycles of quasi-static displacements are performed: the velocity of the intruder during its motion is about 150 µm/s. The upward and downward displacements along the *Y*-axis are separated by rest periods of 9 seconds, during which the system is photographed using a high resolution CCD camera (1280*1024 pixels$^2$) placed above the experimental setup. The camera frame is fixed slightly above the intruder and covers an area of 39 $d_2$*31 $d_2$ (see Figure 1, right): preliminary experiments have shown that the displacements produced by the perturbation are largest above the intruder.

In the following, we use the notation *i* for the index corresponding to the $i^{th}$ image, taken just before the $i^{th}$ displacement of the intruder (upward or downward) and *n* for the cycle number with *n*=int[{*i*+1}/2] where int denotes the integer part. For each image *i*, the centre of each grain is accurately determined using a calculation of the correlation of grey levels between the image of the packing and two reference images corresponding to the two grain types. The inner diameter of the cylindrical hole, which is different for each grain type (small or large), is crucial for the proper determination of the grain type and the position of the particle centre. In the frame of the camera, the diameter of a small grain is $d_1 \sim 26$ pixels and the corresponding value for a large grain is $d_2 \sim 33$ pixels. The above method yields, for each image, the positions of more than a thousand grains with a resolution of 0.05 pixels. The displacement of each grain is then calculated by taking the difference between its position in images *i* and *j*: since the particle displacements are much smaller than a grain diameter, the positions of a given grain in successive images *i* and *j* are identified by locating in image *j* the grain nearest to this grain in image *i*. This method yields a precision of less than 10 µm ($d_2/600$) for the displacements in a given experiment. The displacements have been calculated for 16 independent realizations prepared in the same way. We obtain two types of information:

- The particle displacements in response to an upward or downward intruder displacement (which we refer to as the response function), obtained by comparing two subsequent images *i* and *i+1*. When *i* is odd (respectively even), the intruder is located at its lowest (resp. highest) position and the displacement field produced by the upward (resp. downward) motion of the intruder between images *i* and *i+1* will be referred as an "upward (resp. downward) response" for the $n^{th}$ cycle, where *n*=int[{*i*+1}/2].
- The irreversible particle displacements, obtained by comparing images *i* and *i+2* (between two such images, the intruder has gone through a cycle of displacement, but some grains do not return back to original positions they had before the cycle, i.e., they have undergone irreversible displacements).

**Experimental results and discussion**

*Displacement fields in response to the displacement of the intruder*

In Figure 2a, we present the displacements of grains (magnified by a factor of 50) induced by the second upward displacement of the intruder (*i* =3; *n* =2), i.e, the displacements are the differences in positions of grains between images *i* = 3 and *i* = 4. We observe that the amplitudes of displacements are indeed very small. The displacement of grains is not limited to the vicinity of the intruder: the perturbation due to the intruder has a long range effect, with both vertical and horizontal components. Some recirculation can be observed near the intruder, on both sides.

The displacement fields induced by the subsequent downward displacement of the intruder (*i* = 4; *n* = 2) is shown in (Figure 2b). For the first 30 cycles, the response to the downward displacement essentially consists of a downward displacement of the grains towards the intruder. Due to the bias introduced by gravity these downward displacements are generally larger than the previous upward displacements of the corresponding grains. However, in the vicinity of the intruder, some upward displacements can be observed by an appropriate zoom. These correspond to reversed recirculation around the intruder.

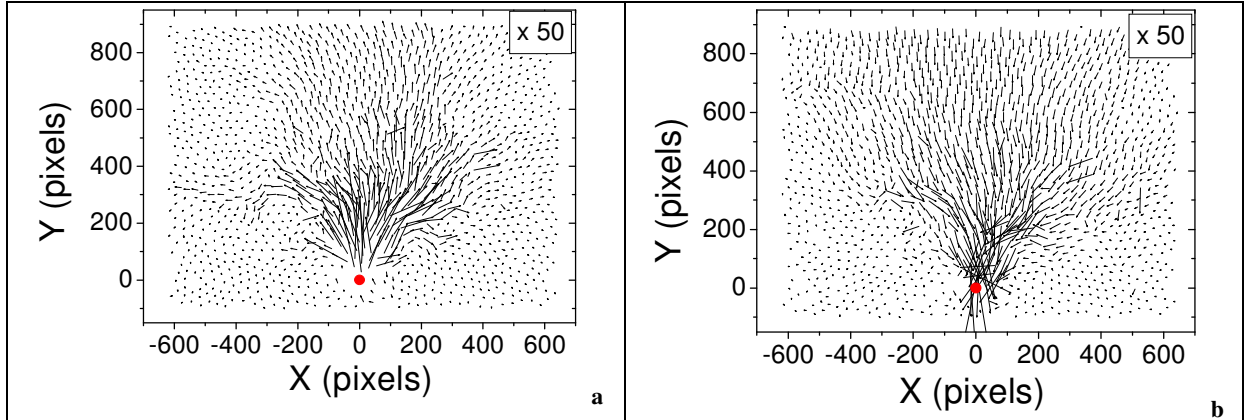

*Figure 2.* Particle displacements observed for a) the second upward displacement ($i = 3$; $n = 2$) and for b) the second downward displacement ($i = 4$; $n = 2$) of the intruder. The red circle corresponds to the position of the intruder. All displacements are magnified by a factor of 50 and the scales are given in pixels. The intruder is initially located at $X = 0$, $Y = 0$ before its upward displacement and comes back to this position after its downward displacement.

As mentioned above, the difference between the upward and downward responses implies the existence of an irreversible field, which can be computed directly by comparing images $i$ and $i+2$ (Figure 3a). This irreversibility leads to a change of the structure of the packing, which modifies the following upward response. We therefore follow the evolution of these responses with the number of cycles. As expected from Figures 2a and 2b, the irreversible field is mainly directed downward corresponding to a very small compaction towards the intruder. The amplitude of the irreversible field typically decays with $i$, which corresponds to a slowing down of the compaction as the system becomes increasingly jammed. In the last stages we need to calculate the irreversible displacements accumulated over several cycles in order to better visualize them. Figure 3b presents the total irreversible field between images $i = 129$ and $i = 515$. The most notable features are the presence of vortices (which are more noticeable during the late stages of the experiment), as well as spatially correlated streams of particles. Note that these types of structures are also observed in the non-affine part of the velocity field in shear simulations of 2D granular material [24] as well as in simulations of plastic deformation in 2D Lennard-Jones amorphous solids [25].

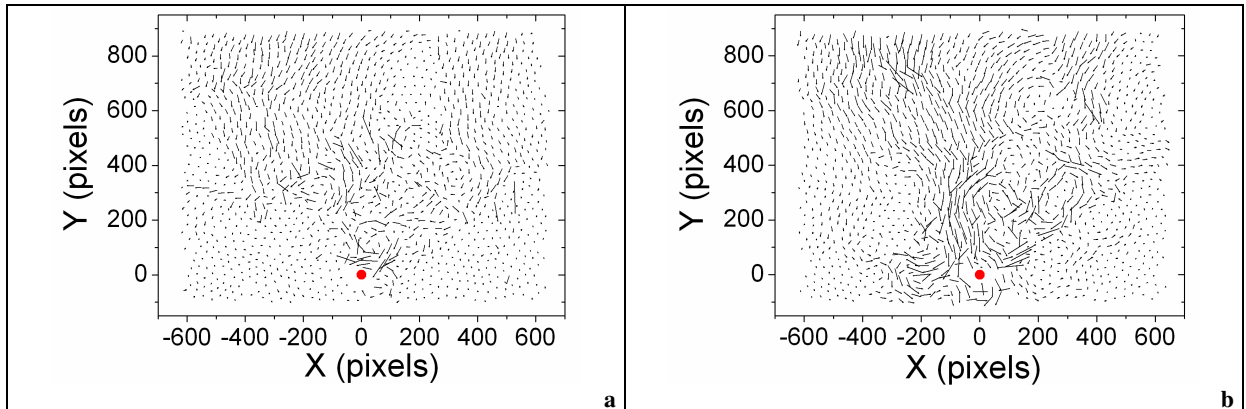

*Figure 3a*: Irreversible particle displacements in the first cycles (shown here for cycle $n = 2$, between images $i = 3$ and $i = 5$), magnified by a factor of 70.

*Figure 3b*: Cumulative irreversible particle displacements between images $i = 129$ and $i = 515$, magnified by a factor of 10. Some vortices are clearly visible.

*Mean upward displacement field*

In order to extract mean displacement field from the particle displacements presented above, we first use the following method [13, 19]: the particle displacements are averaged in small non-overlapping rectangular binning cells of size $1.2d_2*1.2d_2$, covering the entire camera frame, and parallel to the Cartesian axes; they are further averaged over 16 different realizations with the same preparation. This obtained displacement field is presented in Figure 4a for the second upward displacement of the intruder ($i = 3$; $n = 2$). As already observed for the individual particle displacement (Figure 2a), it is clear that the granular displacement is not localized in the vicinity of the intruder, and that this small perturbation (a fourth of a large grain diameter) does indeed induce a far field effect. Furthermore, the presence of two displacement rolls in opposite directions is clearly observed near the intruder, located symmetrically on each side.

The typical decay of the response with the radial distance $r$ from the intruder is analyzed next. We focus here on the asymptotic regime that is obtained after several cycles. We have shown in a previous work [13] that the response is largest above the intruder, with displacement vectors that tend to align with the radial direction from the intruder. Furthermore, we observed that the amplitude of the response to an upward perturbation along the axis of perturbation, i.e., at a polar angle $\theta = 0$ ($\theta$ is defined in Fig. 1b) exhibits a power law dependence, $1/r^\alpha$, where $\alpha$ is close to 1 for sufficiently large $i$. More generally, the upward response (i.e., the displacement induced by moving the intruder up from its reference position at the origin) can be modeled by the following relation:

$$\vec{u}_i^\uparrow \approx b_i U_0 d_2 \frac{f(\theta)}{r} \vec{e}_r$$

This relation is valid in the upper part of the frame, for a distance above the intruder larger than $7d_2$ (sufficiently far from the recirculation rolls). The function $f(\theta)$ of the polar angle $\theta$ has a typical bell shape with a maximum value at $\theta = 0$ (see reference [13] for further details), which corresponds to the direction of the displacement of the intruder. The dimensionless parameter $b_i$ defines the dimensionless amplitude of the response for the $i^{th}$ intruder displacement.

*A more systematic analysis: spatial coarse-grained*

The progressive evolution of the packing can be studied by calculating the changes in the response of the system with the number of cycles. This has first been done by using the binning method described above (see also [19]). However, a more detailed analysis of the displacement field calls for a systematic, resolution-dependent method of calculating it from the measured displacements of the individual particles. In [7], such a method was suggested, based on a spatial averaging (coarse-graining, CG) technique from which the dynamic equations of continuum mechanics can be derived. The displacement, as in continuum mechanics, is defined as the (Lagrangian) integral of the velocity field. The velocity at position r and time t is defined in the usual way, as the ratio of the CG momentum and the CG mass density:

$$\mathbf{V}(\mathbf{r},t) \equiv \frac{\mathbf{p}(\mathbf{r},t)}{\rho(\mathbf{r},t)} = \frac{\sum_k m_k \mathbf{v}_k(t) \phi[\mathbf{r} - \mathbf{r}_k(t)]}{\sum_l m_l \phi[\mathbf{r} - \mathbf{r}_l(t)]}, \quad (1)$$

where $m_k$, $\mathbf{r}_k(t)$ and $\mathbf{v}_k(t)$ are the mass, position and velocity of particle k, and $\varphi[\mathbf{r}]$ is a scalar, non-negative normalized CG function, which possesses a single maximum at $\mathbf{r}=0$, and has a well-defined width $w$. The choice employed here is a Gaussian,

$$\phi(\mathbf{r}) = \frac{1}{\pi w^2} e^{-|\mathbf{r}|^2/w^2}. \quad (2)$$

For the calculations presented in this paper, we use an approximate expression for the displacement [7]:

$$\mathbf{U}(\mathbf{r},t) \approx \frac{\sum_k m_k \mathbf{u}_k(t) \phi[\mathbf{r} - \mathbf{r}_k(0)]}{\sum_l m_l \phi[\mathbf{r} - \mathbf{r}_l(0)]}, \quad (3)$$

valid to leading (linear) order in the strain, where $\mathbf{u}_k(t) \equiv \mathbf{r}_k(t) - \mathbf{r}_k(0)$ is the total displacement of particle. The strain is indeed small in the experiments considered here, the particle displacements being much smaller than their diameter.

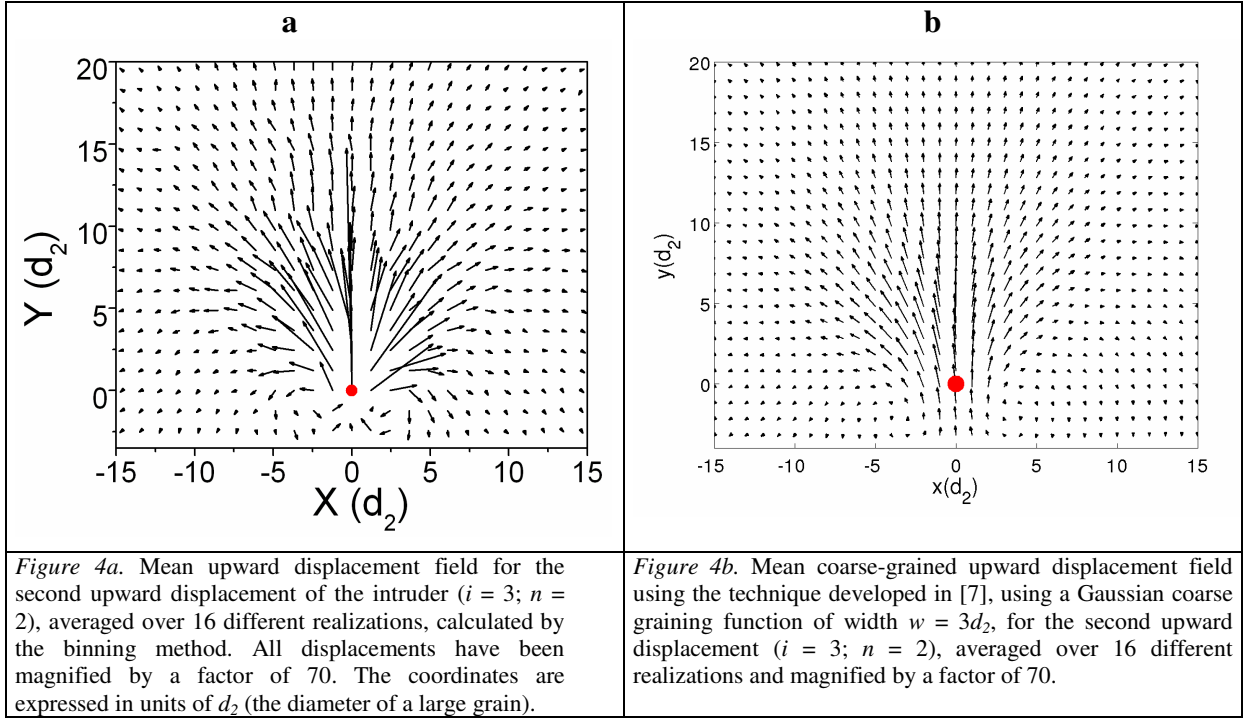

*Figure 4a.* Mean upward displacement field for the second upward displacement of the intruder ($i = 3$; $n = 2$), averaged over 16 different realizations, calculated by the binning method. All displacements have been magnified by a factor of 70. The coordinates are expressed in units of $d_2$ (the diameter of a large grain).

*Figure 4b.* Mean coarse-grained upward displacement field using the technique developed in [7], using a Gaussian coarse graining function of width $w = 3d_2$, for the second upward displacement ($i = 3$; $n = 2$), averaged over 16 different realizations and magnified by a factor of 70.

Figure 4b presents the CG displacement field calculated using a CG width $w = 3d_2$. As for the binning method, the displacement field has been averaged over the ensemble of $N = 16$ realizations with the same preparation. Since the displacement gradients in the system (at least close to the intruder) are quite large on the particle scale, the displacement depends on the CG scale [20], and it is hard to identify a scale-independent regime in the displacement, even for the ensemble average. The choice $w = 3d_2$ corresponds to the range of CG widths with the weakest scale dependence (i.e., the smallest derivative with respect to $w$). Note that the CG method used here does not account specifically for the presence of boundaries, outside of which there are no particles. In the context of the present paper, we note that such boundaries effectively occur at the boundaries of the camera frame, as well as around the intruder (there are difficulties in detecting the particles in a radius of $2d_2$ around to it due to shadows cast by the experimental apparatus used for moving the intruder). Therefore, the field values calculated within a margin of about $w$ from the boundary and $2d_2+w$ around the intruder may not be reliable. Even if we obtain the same features for the displacement field as in Figure 4a, one advantage of this CG method is that it enables the calculation of a smooth displacement field: unlike in the binning method, the field is defined at any point in the plane, and is fully consistent with continuum mechanics. This also enables the calculation of all the displacement gradients [7], e.g. the strain components. **In the following, we apply this CG method together with ensemble averaging**.

*Evolution of the response with the number of cycles*

In order to characterize the evolution of the packing quantitatively, we calculated the (ensemble averaged) radial displacement at $X = 0$ along the vertical direction (along $\theta = 0$) for the different values of $i$.

These displacements (Figure 5, green symbols for the "normal" preparation described before) can be fit by a power law in the distance $r$ from the intruder:

$$u_r(r, \theta = 0) = A / r^\alpha.$$

This formula expression for the fitting function is still valid in the last stages of the experiment and corroborates the result obtained with the binning method for the asymptotic response (large $i$), for which $\alpha \approx 1$ [13]. Note that we restrict the fit to $r > 7d_2$; the maximum value of $r$ in the fit is about $21\ d_2$, due to the margin of order $w$ from the boundary of the camera frame for which the displacement field calculation may be unreliable, as mentioned above. As the amplitude of the displacement field decays with $r$, it is likely that the system evolves less when far from the intruder. However, the power law fit characterizes the overall response of the packing over the entire range of $r$ values, and therefore cannot capture local effects directly: the parameters $A$ and $\alpha$ of the fit can only provide a global characterization of the evolution of the packing. Although the range of $r$ involved in the fit is less than a decade (so that the power law behavior should not be taken very seriously), the obtained parameters $A$ and $\alpha$ show a clear evolution with the number of cycles, as shown in Figure 6.

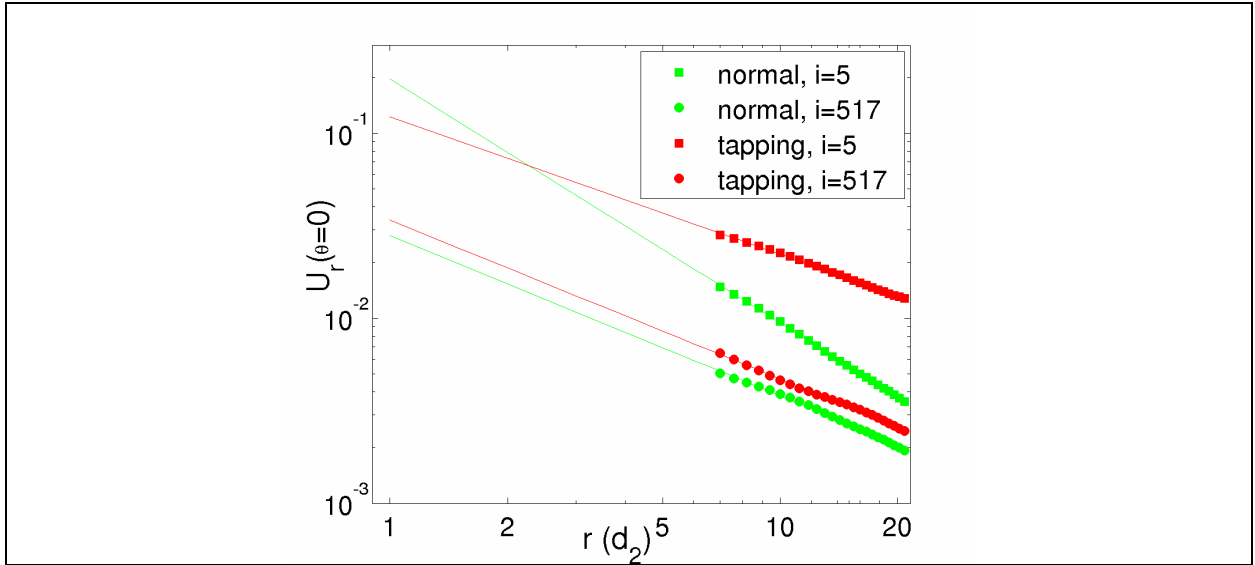

*Figure 5*: Radial component of the ensemble averaged CG displacement along the vertical direction above the intruder as a function of the distance from the intruder. The green symbols (resp. red) correspond to the normal preparation (resp. tapping). The lines are power law fits to the data. Note that the scales for both axes are logarithmic.

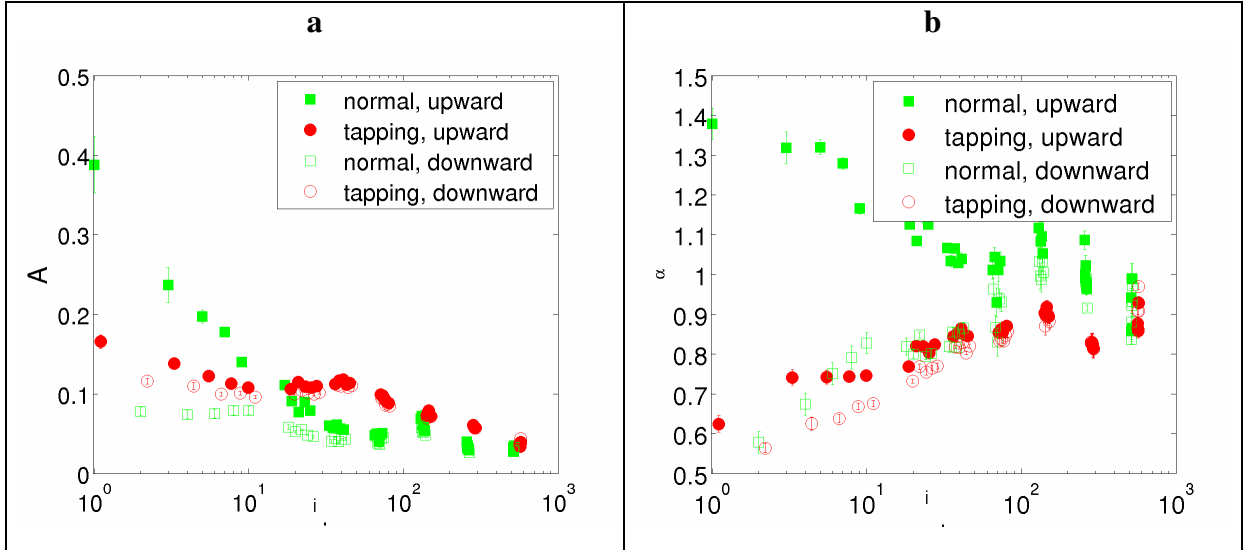

*Figure 6:* The parameters $A$ and $\alpha$ of the power law fit $u_r(r, \theta = 0) = A/r^{\alpha}$ to the CG displacement field at $X = 0$ as a function of the number $i$ of displacements of the intruder. The green square symbols (resp. red point symbols) correspond to the normal preparation (resp. tapping). The abscissas for the two preparation methods are slightly shifted, for clarity.

The error bars shown in Figure 6 arise from the fitting procedure, and do not take into account the fluctuations in the displacement field within the ensemble. These fluctuations are rather large (the standard deviation and the mean are of the same order of magnitude). If we take the standard deviations of the fluctuations as "errors" in the displacement field and estimate the errors in the fit parameters using a weighted fit [23], it yields errors in the parameters which are larger by several orders of magnitude.

For the preparation described above ("normal", green squares in Figures 6a and 6b) the parameter $A$ decreases progressively with $i$ and then saturates to a constant value after about 60 displacements, i.e. 30 cycles. Simultaneously, the exponent $\alpha$ tends to 1, as already obtained using the binning method [13,19]. After $n \approx 30$ cycles, the absolute values of the mean response for upward and downward displacements of the intruder are almost identical: a quasi-stationary regime or "limit cycle" is obtained (i.e., the mean displacement response becomes nearly reversible) and the packing seems to be more stable with respect to the perturbation. The parameters corresponding to the downward displacement [obtained from a fit to $-u_r(r, \theta = 0)$] are shown as well; the fact that the irreversible part decays with $i$ is reflected in the convergence of $A$ and $\alpha$ for the upward and downward displacement.

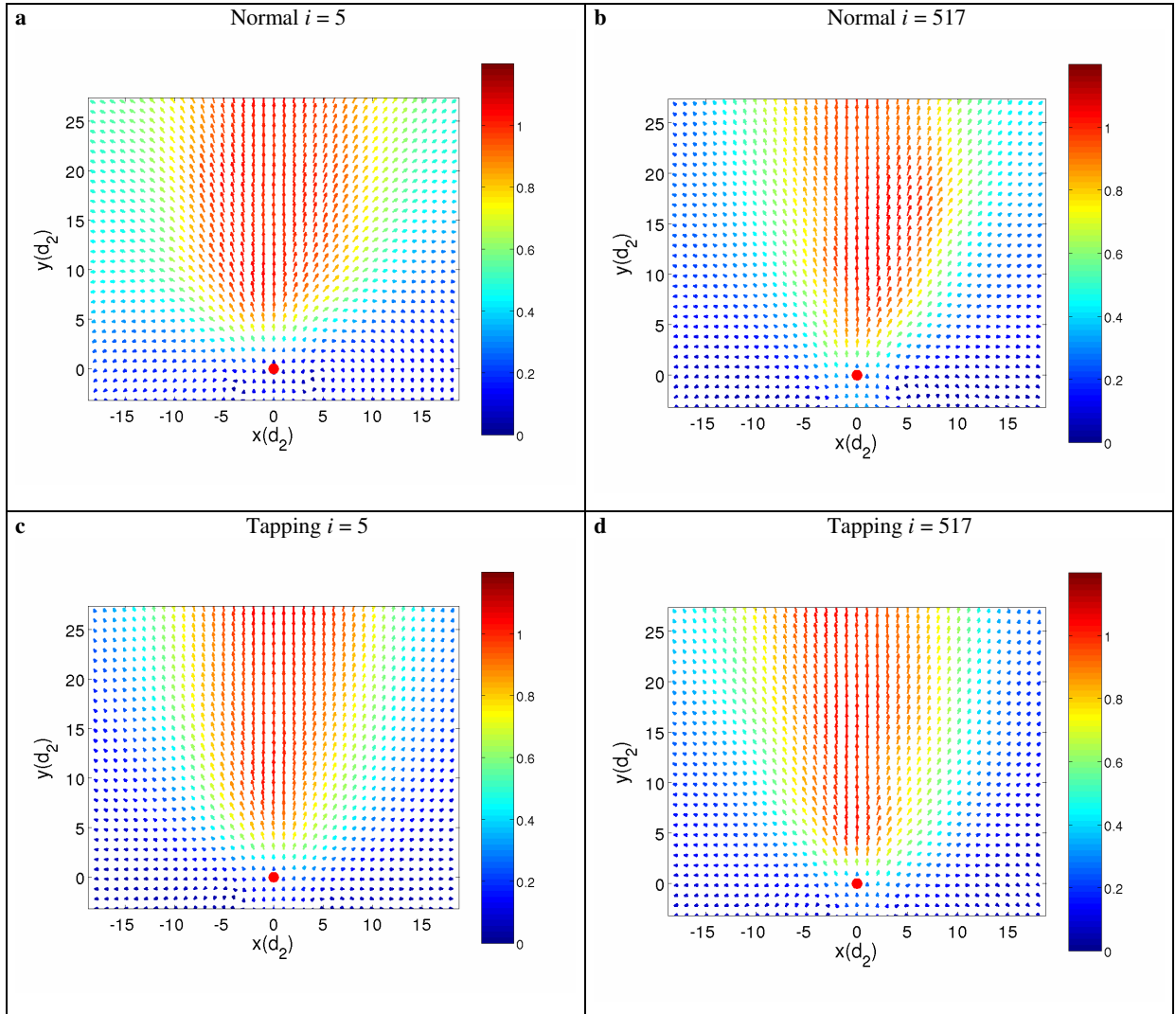

*Figure 7*: The upward CG displacement field for the two preparation methods, in cycles $n = 3$ ($i = 5$) and $n = 254$ ($i = 517$): We use a Gaussian CG with $w = 3\ d_2$, averaged over 16 experiments for each preparation. The field is rescaled by multiplying it by $r^\alpha / A$, where $A$ and $\alpha$ are the parameters obtained from a power law fit at $X = 0$ (see text). The color indicates the magnitude of the rescaled field.

Figures 7a and 7b show the CG upward displacement field at the early stages ($i = 5$) and towards the end of the experiment ($i = 517$) for the "normal" preparation. As the response decays with the distance from the intruder and with the number of cycles $n$, we rescaled all the CG fields, multiplying by $r^\alpha / A$, where $A$ and $\alpha$ are the fit parameters obtained before, for the corresponding displacement $i$. This rescaling enables an easier visualization and comparison of the relative lateral decay of the response away from the vertical axis. It also shows the direction of the CG field. The color code is based on the same rescaling. Note that the rescaled amplitude of the displacement should be of the order of 1 along the vertical direction, where the fit was performed. A comparison of Figures 7a and 7b clearly shows a smaller lateral displacement (with respect to the displacement along the vertical axis at $x=0$) at the last stages: the amplitude of the rescaled field is smaller closer to the vertical direction (see the color code) and the arrows tend to better align with the radial direction for $i = 517$.

Although the evolution of the mean response is quite evident from the evolution of $A$ and $\alpha$, there is almost no change in the mean packing fraction (averaged over the 16 experiments) with the number of cycles $n$: a very small compaction produced by the repeated displacement of the intruder is revealed through the existence of a downward irreversible field (Figure 3) but the corresponding increase of the packing fraction is less than 0.1 % over the 261 cycles (see Figure 8).

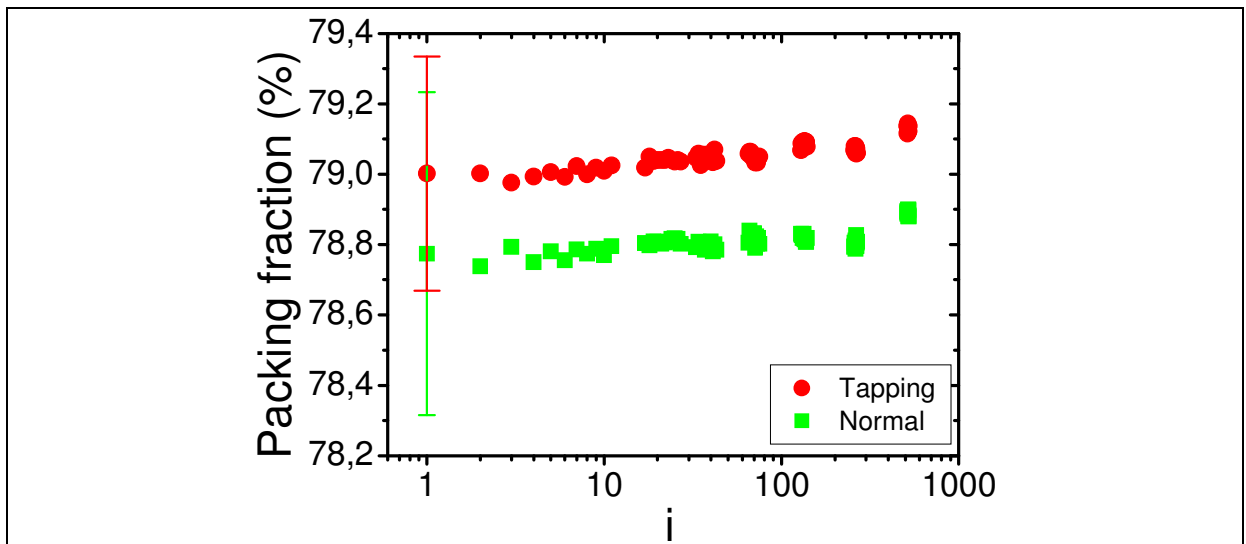

*Figure 8:* Packing fraction averaged over the 16 realizations, as a function of the number $i$ of displacements of the intruder. The indicated error bars are the standard deviations over the 16 realizations. Note that the errors are larger for the normal preparation.

*Dependence on preparation*

We now compare two methods of preparation described above: the "normal" and "tapping" method. As mentioned, the difference in preparation does not change the packing fraction significantly: $\Delta c / c = +0.3$ % (Figure 8); however, the effect of the preparation on the subsequent response, characterized by the mean CG displacement field, is quite significant, as shown in Figures 5 and 6 and by comparing Figures 7a and 7c. The difference in the observed response between the preparation methods is quite pronounced in the first cycles. For the tapping preparation, the overall displacements are larger: compare for example the green and red squares in the plot of $u_r(r, \theta = 0)$ as a function of $r$ in Figure 5.

For the "tapping" preparation, the evolution of the amplitude of the displacement along the direction $\theta = 0$ can also be fit by a power law, $u_r(r, \theta = 0) = A / r^\alpha$. As for normal preparation, the amplitude $A$ decays with the number of cycles (red points in Figure 6a), but seems to undergo a smaller change. The exponent $\alpha$ (for the upward displacement) exhibits a very different behavior for the two preparation methods: while it decreases from 1.4 to 1 for the normal preparation (full green squares in Figure 6b), it increases from 0.6 to 1 for the tapping preparation (full red points in Figure 6b). This increase of $\alpha$ indicates a progressive decrease in the range of the perturbation along the vertical axis, and is probably associated with the fact the response becomes radial (or more isotropic) with the number of cycles. In both cases, the field becomes more radial. The difference in the downward displacement between the preparations is much smaller (in particular, $\alpha<1$ for the downward displacement obtained with both preparation methods), which appears to support the notion that the downward displacement is dominated by the effect of gravity rather than by the structure of the packing. Both $A$ and $\alpha$ for the two preparation methods approach each other as the system evolves. More generally, both the magnitude and direction of the mean displacement field in the two preparations become closer (compare Figures 7b and d). We note that other experiments [21] as well as simulations [22] have demonstrated that the stress response is also significantly influenced by the method of preparation.

*Coordination number*

As mentioned above, the packing fraction appears to be an insufficiently sensitive parameter for discriminating between different preparation methods. Furthermore, in both cases, the change in the mean packing fraction during the experiment is less than 0.1 %. It is then natural to look for a more microscopic characterization. To this end, we consider the coordination number $Z$, i.e., the number of contacts per grain.

Once the positions of the centres of grains and their respective types (large or small grain) are known, it is possible in principle to determine which grains are in contact. However, in the real experiment there is a slight polydispersity in the diameters of each type of grains and the positions of the centres are only determined with a finite precision (0.05 pixel / 26 pixels corresponding to ~ 0.2% of the diameter of a small grain). The main error comes from the distribution of diameters; the tolerance of the outer diameters (given by the manufacturer of the grains) is -70 μm (~ - 0.5 pixels) over 4 and 5 mm, yielding a maximum error of the order of 1.75%. Therefore, in order to calculate the coordination number, we introduce a tolerance $\varepsilon$ in the criterion for determining the existence of a contact. Two grains $i$ and $j$ separated by a distance $d_{ij}$ (calculated from the positions of their respective centres as obtained from the image analysis) are estimated to be in contact if:

$$d_{ij} \leq r_i^0 + r_j^0 + \varepsilon$$

where $r_i^0$ and $r_j^0$ are the lower bounds for the radii of the corresponding grains ($r_i^0$ = 26 or 33 pixels for the small and large grains, respectively).

The choice of the tolerance $\varepsilon$ has a dramatic influence on the mean coordination number <$Z$>, as can be seen in Figure 9a. <$Z$> has been calculated in the same window as the packing fraction, namely in a rectangular sub-frame at a distance of 3 $d_2$ above the intruder and at 1.5 $d_2$ from the left, top and right boundaries of the camera frame. <$Z$> is averaged over this window and over the ensemble of 16 realisations with the same preparation, for the first image. When $\varepsilon \rightarrow 0$, the fact that <$Z$> $\rightarrow$ 0 simply means the absence of detected contacts since the radii $r_i^0$ or $r_j^0$ are smaller than the actual radii of the grains. Increasing $\varepsilon$ is equivalent to increasing the effective radii of grains, so that the grains progressively touch each other. This leads to a rapid increase of <$Z$> from 0 to 4 over a small increase in $\varepsilon$ (less than a pixel). For larger $\varepsilon$, we observe a slower increase of <$Z$> with $\varepsilon$. Obviously, if $\varepsilon$ is too large, false contacts are detected (since the effective radii become unrealistically large).

In order to choose the appropriate value of $\varepsilon$ for discriminating between real and false contacts, we analyse the populations $nZx$ of grains with coordination number $x$ as a function of $\varepsilon$ (e.g., $nZ2$ is the total number of grains having two contacts, summed over the 16 realisations), presented in Figure 9b. Note that the total number of grains considered over these realizations is 16362, which provides rather good statistics. We observe a gradual decrease of $nZ0$ with $\varepsilon$, while the populations of larger coordination numbers increase at increasingly larger values of $\varepsilon$, and then decrease. Our choice of $\varepsilon$ is based on the following physical argument: in real packings lying on an inclined plane, the probability of having grains with only one contact should be rather small. We therefore choose a value of $\varepsilon$ for which the population $nZ1$ nearly vanishes; specifically, we chose $nZ1$ equal to 0.5 % of the total population of grains, rather than zero, due to the finite friction between grain and bottom plate as well as among the grains. The corresponding value of $\varepsilon$ is $\varepsilon_0$ =1.722 pixels, which can be considered as the upper bound required for scanning the full distribution of real diameters.

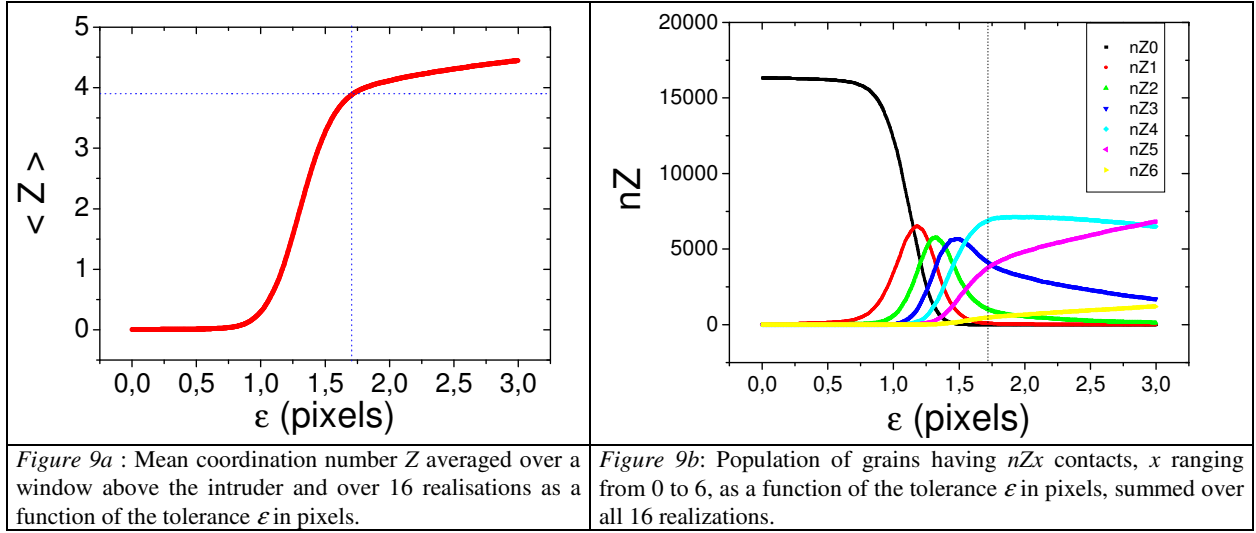

*Figure 9a*: Mean coordination number Z averaged over a window above the intruder and over 16 realisations as a function of the tolerance $\varepsilon$ in pixels.

*Figure 9b*: Population of grains having $nZx$ contacts, $x$ ranging from 0 to 6, as a function of the tolerance $\varepsilon$ in pixels, summed over all 16 realizations.

Using the value $\varepsilon_0$, we recalculate the coordination number for all images of the 16 experiments and for both preparation methods. Figure 10a shows the resulting mean coordination number as a function of the image number. As expected, the tapping preparation leads to a slightly higher coordination number ($<Z> = 3.98 \pm 0.04$) than the normal preparation, for which $<Z> = 3.90 \pm 0.05$. The error bar is the standard deviation in the ensemble. This increase in $<Z>$ is associated with a larger packing fraction (see Figure 10b). However, the relative increase (for the tapping preparation) in packing fraction is only $\Delta c / c = +0.3$ %, while the relative increase in the coordination number, $\Delta<Z>/<Z>$, is +2.0 %. Thus $<Z>$ is a more sensitive parameter for characterizing the differences between the two preparation methods. This increase in $<Z>$ is associated with a relative decrease (in proportion of the total number of grains) in the populations $nZ2$ and $nZ3$, in favour of $nZ5$ and $nZ6$. Grains with two contacting neighbors may be considered to be situated beneath a local arch (i.e., they have no contacting neighbors above them). We therefore suggest that the decrease in $nZ2$ and $nZ3$ is probably related to the breaking of local arches due to the tapping preparation. A reduction of the number of these arches should also yield a packing in which a smaller part of the stress is transmitted towards the lateral walls, thus increasing the displacement along the axis of the perturbation, as indeed observed in Figure 5 and in Figure 7c compared with Figure 7a, especially just above the intruder. For the normal preparation, the presence of these arches leads to a broader response and to a more rapid decay of the amplitude of displacement in the axis of the perturbation (as exhibited by the faster decay, $\alpha > 1$, for the earlier cycles; see Figure 6b).

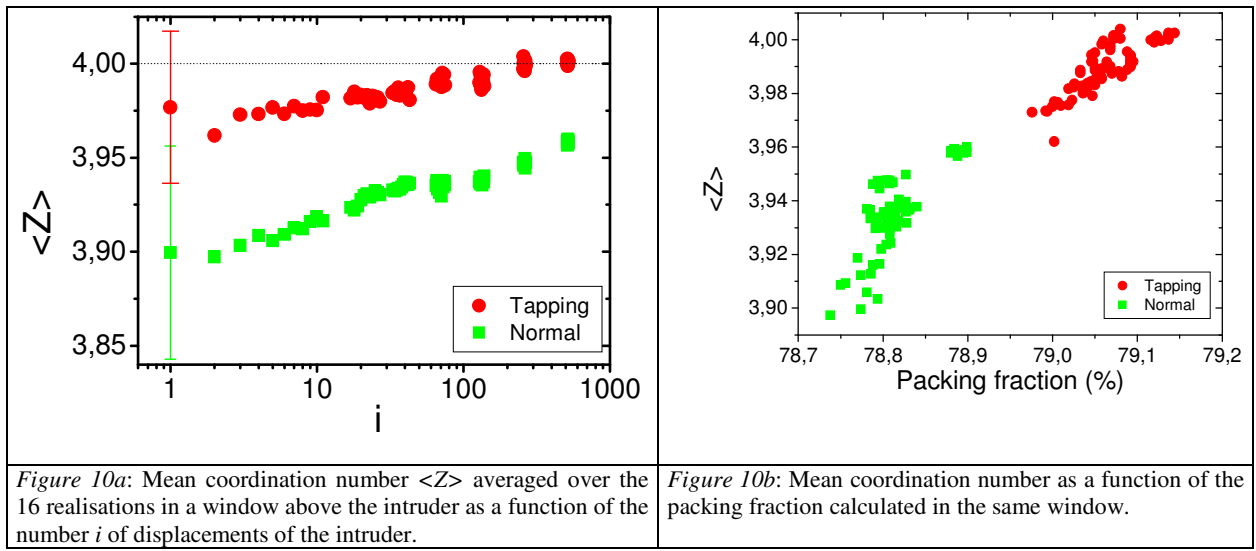

*Figure 10a*: Mean coordination number $<Z>$ averaged over the 16 realisations in a window above the intruder as a function of the number $i$ of displacements of the intruder.

*Figure 10b*: Mean coordination number as a function of the packing fraction calculated in the same window.

The further evolution with the number of cycles shows a quasi-logarithmic increase of the coordination number (which is again more noticeable than the increase in the packing fraction $c$ with $i$ - see Figure 8), which depends on the initial preparation. For the normal preparation, the population of $nZ2$ decreases with $i$ so that the cyclic perturbation apparently breaks down the arches initially present in the normal preparation (at least sufficiently close to the intruder). As explained above, the different proportion of grains involved in arches might be related to the fine differences observed in the displacement fields. A smaller number of arches (or a more isotropic distribution of contacting grains) might lead to reduced lateral transmission of stress and a slower decay of the displacement along the vertical axis: $\alpha$ decreases with $i$ for the normal preparation, so that the range of the perturbation along the vertical axis increases with $i$. When prepared by tapping, the grains are more jammed, so that rearrangements do not occur as easily as for the normal preparation; consequently, $<Z>$ increases more slowly with the number of cycles (see Figure 10a).

**Conclusion**

The rearrangements due to a small localized cyclic displacement applied to a packing of rigid grains under gravity in a 2D geometry were measured experimentally. They are characterized by different methods, including a direct observation of the particle displacements, the calculation of the mean displacement field (averaged over space as well as over an ensemble of realizations), the packing fraction and the coordination number. Quite surprisingly, we find that the displacement response is rather long ranged, and evolves considerably with the number of cycles. We study the effect of a small difference in the preparation method: initial configurations prepared by random mixing of grains in a given area and by an additional weak tapping clearly exhibit a different response, even though the mean packing fractions obtained in the two cases are extremely close. The preparation affects not only the initial response but also its further evolution during the cycling procedure. It demonstrates that the system still retains a memory of the initial preparation after many cycles. The last stage of the experiment, however, seems to indicate that the mean displacement fields for both preparation methods tend to the same response: a nearly radial field with a $1/r$ decay with the distance from the intruder. The evolution of displacement fields (direction and magnitude) has been tentatively linked to microscopic features inferred from the estimated coordination number. The coordination number is shown to be a more sensitive parameter than the packing fraction for characterizing both the initial configuration and its subsequent evolution. In particular, the presence of a nonuniform redistribution of contacts around a grain (in particular, for grains having a coordination number of two, i.e., belonging to small arches) appears to spread out the response laterally, away from the axial direction. The response to a local perturbation is therefore shown to be a very sensitive probe of the degree of jamming in a granular material.

**Acknowledgements**

We would like to thank I. Goldhirsch and I. Aronson for useful discussions. CG gratefully acknowledges the support of the Chateaubriand Fellowship (French Ministry of Foreign Affairs), the Arc-en-Ciel – Keshet exchange program, the German-Israeli Foundation (GIF), grant no. 795/2003, and the US-Israeli Binational Science Foundation (BSF), grant No. 2004391.